\documentclass[11pt]{article}
\usepackage[margin=1in]{geometry}
\usepackage{amsmath}
\usepackage{amssymb}
\usepackage{booktabs}
\usepackage{graphicx}
\usepackage{hyperref}
\usepackage{natbib}
\usepackage{xcolor}
\usepackage{placeins}
\usepackage{float}
\usepackage{microtype}
\hypersetup{colorlinks=true, linkcolor=blue, citecolor=blue, urlcolor=blue}

\title{\textbf{Orchestra: Corroboration-Based Regulatory Candidate Discovery via Composed Bioinformatics MCP Agents}}
\author{Jose A. Bird \\ \small ORCID: 0009-0006-2744-0606 \\ \small Affiliation: Independent Researcher \\ \small Correspondence: jbird@birdaisolutions.com}
\date{}

\begin{document}
\maketitle

\begin{abstract}
Orchestra composes two independently-built bioinformatics MCP servers -- RegNetAgents \citep{bird2026regnetagents}, which infers regulatory network topology, and CASCADE \citep{bird2026cascade}, which provides independent functional evidence (LINCS knockdown, DepMap essentiality, super-enhancer status, DoRothEA transcription-factor confidence) -- into a single multi-agent LangGraph workflow \citep{langgraph} exposed via the Model Context Protocol \citep{mcp}. Orchestra's central architectural claim is that requiring RegNetAgents' network-topology evidence and CASCADE's independent functional evidence to \emph{agree} on a candidate regulator yields a more trustworthy candidate than either system alone. This claim has not previously been tested directly: RegNetAgents' own validation asked only whether its candidate lists beat chance, never whether cross-system corroboration adds precision beyond either system individually. We test it on the TCGA \emph{tumor-acquired regulator tier} -- regulators present in a gene's tumor-state ARACNe network but absent from GREmLN's population-averaged baseline (not from matched normal tissue) -- the tier in which RegNetAgents' own validation found a modest excess of known cancer driver genes over random genes.

Candidate selection matters to this test, so every experiment here selects the ten regulators with the highest ARACNe mutual-information (MI) edge weight, RegNetAgents' own per-edge network-confidence signal, rather than an arbitrary subset. Across RegNetAgents' BRCA/COAD focal-gene panel plus matched negative-control panels, requiring genuine multi-source agreement ($\geq 2$ of 4 independent CASCADE evidence sources) predicts known-driver status (IntOGen) among focal genes (corroborated vs.\ uncorroborated candidates, odds ratio $=3.15$, Benjamini--Hochberg-adjusted $p=0.0162$; $2.92$ after deduplication) but shows no association among negative controls tested identically ($p=0.70$). In BRCA/COAD, a single evidence source ($\geq 1$) predicts driver status in neither focal genes nor negative controls, so the $\geq 2$ effect reflects genuine multi-source agreement rather than the mere presence of any evidence. Two is the smallest count at which independent sources can agree at all, so we adopt it as the operational definition of agreement; higher thresholds are not testable at current sample sizes.

We further test this pattern in a third cancer type, STAD, on a second, independently constructed panel whose regulator pools are drawn from a separate patient cohort: the pattern reproduces and is stronger (OR $=6.71$, Benjamini--Hochberg-adjusted $p=0.0047$; $6.40$ after deduplication), while negative controls again show no association ($p=0.66$). Single-source evidence is not stable across panels -- null in both BRCA/COAD groups, but in STAD significant in focal genes after Benjamini--Hochberg correction and not in negative controls. Requiring $\geq 2$ sources, by contrast, separates focal genes from negative controls in both panels.

MI edge weight is a strong single predictor in BRCA/COAD ($p=0.0008$, a lower $p$-value than any individual CASCADE source reaches there) but carries no detectable signal in STAD ($p=0.50$), so we directly tested whether corroboration adds value beyond it specifically: it does, in both panels (BRCA/COAD $p=0.0386$; STAD $p=0.0001$) -- the most direct evidence here that cross-system agreement captures signal neither system supplies alone, though modest in BRCA/COAD. This paper reports four robustness checks applied to the core $\geq 2$ finding: deduplication of repeated candidates, a reproduction cross-check against RegNetAgents' own published results, a permutation test that builds its null distribution directly from the data, and a check against a second, separately derived driver list (Vogelstein et al. 2013) -- which points the same way but does not reach significance in either panel and is too sparse to confirm the pattern on its own. Every experiment in this paper invokes \texttt{OrchestraWorkflow.run\_analysis()} directly, Orchestra's real agentic entry point, mirroring the validation methodology of \citet{bird2026cascade}.
\end{abstract}

\section{Introduction}
\label{sec:intro}

It is increasingly common to build multi-agent AI systems by connecting separately-built agents through a shared communication protocol -- such as the Model Context Protocol \citep{mcp} -- rather than merging their underlying data and analysis logic into one system. The implicit claim behind any such composition is that when independent agents agree, that agreement is more trustworthy than either alone -- the same logic behind aggregating independent estimators to reduce error. But that logic only holds if the agents' errors are genuinely uncorrelated, and whether any \emph{particular} composed system actually satisfies that condition is rarely tested directly: two agents could agree for reasons that have nothing to do with correctness -- overlapping training data, correlated heuristics, or a shared confound -- for instance, both systems tending to favor genes that are simply well-studied in general, regardless of whether that makes them relevant to the question at hand. Whether that condition actually holds for any given composed system is an empirical question, not something a general argument for ensembling can settle in advance. This paper tests it directly, for one composed system, rather than assuming the answer.

RegNetAgents, CASCADE, and Orchestra are built by the same author, and as of this writing none has undergone independent peer review. Because the paper's central claim turns on the two evidence systems being independent, shared authorship is a genuine threat to it -- one the negative-control panels (Section~\ref{sec:panel}) address directly, and the second driver list (Section~\ref{sec:cgc-check}) probes, though it is too sparse to be conclusive.

Orchestra composes RegNetAgents, which infers gene regulatory network topology from ARACNe co-expression networks \citep{margolin2006aracne, lachmann2016aracne}, and CASCADE, a perturbation-analysis engine that simulates downstream regulatory effects via network propagation -- a capability previously validated against real patient tumor data \citep{bird2026cascade} -- and which also provides four lightweight, independently-checkable evidence sources used for corroboration in this paper: LINCS L1000 knockdown perturbation profiles \citep{subramanian2017lincs}, DepMap CRISPR essentiality screens \citep{tsherniak2017depmap}, curated super-enhancer annotations \citep{khan2016dbsuper, loven2013superenhancers}, and DoRothEA transcription-factor confidence tiers \citep{garciaalonso2019dorothea}. These two sub-agents are a good test case for the independence question above precisely because their evidence is methodologically distinct by construction: one is inferred from co-expression statistics across a reference corpus, the other drawn from perturbation and dependency assays and curated functional annotations -- not two views of the same underlying dataset. RegNetAgents' own validation \citep{bird2026regnetagents} tested whether its regulator candidate lists, classified by which of two ARACNe network sources they appear in (a population-averaged single-cell atlas from the GREmLN project \citep{zhang2025gremln} versus TCGA tumor-state bulk networks \citep{aracne_networks, giorgi2026aracne}), are enriched for known cancer driver genes relative to random genes: tumor-network candidates showed a modest excess (consistent in COAD, borderline in BRCA), and population-averaged candidates showed none. That validation asks whether a candidate list is better than chance; it does not ask whether \emph{agreement} between RegNetAgents' network-topology evidence and CASCADE's independent functional evidence is itself a meaningful filter, which is Orchestra's specific architectural contribution and the question this paper addresses. Beyond this synthetic validation, Orchestra has also supported biological discovery directly: a companion short communication built on Orchestra's synthesis outputs, analyzing DACH1-methylation correlations in TCGA cervical squamous carcinoma, is available as a posted preprint \citep{bird2026dach1}.

This is not a biological discovery paper: it reports no new regulator--gene relationship validated by experiment. It is a validation report for one Orchestra feature -- cross-system corroboration scoring, exercised through Orchestra's real agentic entry point rather than a bypass script -- and, through that feature, a direct test of the independence question raised above.

\subsection{Contribution}

\begin{enumerate}
\item A direct empirical test of Orchestra's cross-system corroboration claim on the TCGA tumor-acquired regulator tier -- the candidate population in which RegNetAgents' own paper found a modest excess of known cancer driver genes over random genes.
\item A demonstration that requiring genuine multi-source agreement ($\geq 2$) separates focal genes from negative controls in both panels, whereas a single evidence source ($\geq 1$) is not stable across panels: in BRCA/COAD $\geq 1$ reaches significance in neither group at any correction level; in STAD it reaches significance in focal genes only (adjusted $p=0.0083$; Section~\ref{sec:stad-results}).
\item A principled candidate-selection mechanism: ordering each query's tumor-acquired regulators by ARACNe MI edge weight before the 10-candidate cap, rather than accepting RegNetAgents' arbitrary alphabetical output order -- resolving an open limitation in Orchestra's original candidate selection.
\item Four robustness checks on the $\geq 2$ result under this selection: (i)~deduplication of repeated candidates, (ii)~a reproduction cross-check against RegNetAgents' own published results, (iii)~a 1,000-draw permutation test that builds its null distribution directly from the data, and (iv)~a second, separately derived driver list (Vogelstein et al. 2013), which is directionally consistent but too sparse to reach significance.
\item Reproduction in a second, separately constructed panel -- STAD, a third cancer type -- where the $\geq 2$ pattern reproduces and is stronger (deduplicated OR $=6.40$, vs.\ $2.92$ for BRCA/COAD).
\item An opt-in, backward-compatible implementation (\texttt{validate\_tumor\_acquired}, with a second opt-in parameter \texttt{rank\_tumor\_acquired} controlling candidate selection) in Orchestra's production \texttt{compare\_network\_contexts} tool, exercised directly by every experiment in this paper via \texttt{OrchestraWorkflow.run\_analysis()} rather than a bypass script, following the same real-agentic-entry-point validation principle established by \citet{bird2026cascade}.
\item A direct test of whether cross-system corroboration adds predictive value beyond MI edge weight -- a stronger single predictor than any individual CASCADE source in BRCA/COAD, though not in STAD -- using a logistic-regression likelihood-ratio test. Corroboration adds significant value in both tested panels (BRCA/COAD $p=0.0386$; STAD $p=0.0001$), the most direct evidence in this paper for Orchestra's central architectural claim, though modest in BRCA/COAD.
\end{enumerate}

\section{Methods}

\subsection{Orchestra architecture and the validated code path}
\label{sec:architecture}

Orchestra does not call RegNetAgents and CASCADE as stateless APIs; it orchestrates them as two independent agentic sub-systems, each its own LangGraph-orchestrated workflow \citep{langgraph} with its own published validation \citep{bird2026regnetagents, bird2026cascade}, composed at the Model Context Protocol layer \citep{mcp}. A top-level LangGraph workflow classifies each incoming query's analysis type and routes it to one of Orchestra's eleven composite tools, each encoding a different coordination pattern between the two sub-agents.

Some paths dispatch to both sub-agents \emph{concurrently}: \texttt{causal\_chain\_analysis}, for example, issues one call to RegNetAgents and one to CASCADE simultaneously via \texttt{asyncio.gather}, appropriate when neither call depends on the other's output. This paper's experiments instead exercise \texttt{compare\_network\_contexts} (\texttt{\_run\_network\_comparison\_path}), which requires a \emph{sequential}, dependent pattern: it first calls RegNetAgents' own \texttt{compare\_network\_contexts} tool, which returns three regulator sets for a focal gene and TCGA cancer type -- \textbf{conserved} (present in both the GREmLN population-averaged network and the TCGA tumor-state network), \textbf{population-averaged-only}, and \textbf{tumor-acquired} (present in the TCGA network only) -- and only then decides which of them, if any, to forward to CASCADE for independent validation, since CASCADE's inputs are themselves an output of RegNetAgents' call rather than available up front.

Prior to this work, that step forwarded the conserved regulators to CASCADE for cross-system corroboration but not the tumor-acquired ones -- even though RegNetAgents' own paper found tumor-network candidates to show a modest excess of known cancer driver genes over random genes, and population-averaged candidates none \citep{bird2026regnetagents}.

We added an opt-in boolean parameter, \texttt{validate\_tumor\_acquired} (default \texttt{False}), that extends this path to also validate up to 10 tumor-acquired regulators against the four independent CASCADE evidence sources of Section~\ref{sec:corroboration}. Which 10 are selected, from what is often a much larger pool (Section~\ref{sec:ranking}), is controlled by a second opt-in parameter, \texttt{rank\_tumor\_acquired}: by default the first 10 in RegNetAgents' alphabetical order are used, and setting it to \texttt{True} instead orders the pool by ARACNe MI edge weight before the cap (it has no effect unless \texttt{validate\_tumor\_acquired} is also \texttt{True}). Every experiment in this paper sets both to \texttt{True}; default behaviour for existing callers of \texttt{compare\_network\_contexts} is unchanged, and the full unit-test suite passes with zero regressions.

Every experiment calls Orchestra's real workflow function, \texttt{OrchestraWorkflow.run\_analysis()}, directly -- the same routing, sub-agent calls, and synthesis that a live query runs -- not a separate script that reimplements that logic. This follows the validation approach of \citet{bird2026cascade}: exercise the agentic tool's real code path, not a hand-rolled equivalent.

\subsection{Candidate selection: ranking tumor-acquired regulators by ARACNe MI edge weight}
\label{sec:ranking}

The tumor-acquired pool RegNetAgents' \texttt{compare\_network\_contexts} returns ranges from 8 to 84 candidates per query (averaging 24--28 across this paper's panels), so the 10-candidate cap (Section~\ref{sec:architecture}) drops most of it in almost every query (all STAD queries, 88\% of BRCA/COAD queries). Which 10 are kept therefore matters -- and the pool arrives in alphabetical order by gene symbol (the \texttt{tumor\_state\_only} field), an ordering that carries no information about confidence or importance, so keeping the first 10 would be arbitrary.

\begin{figure}[H]
\centering
\includegraphics[width=0.85\textwidth]{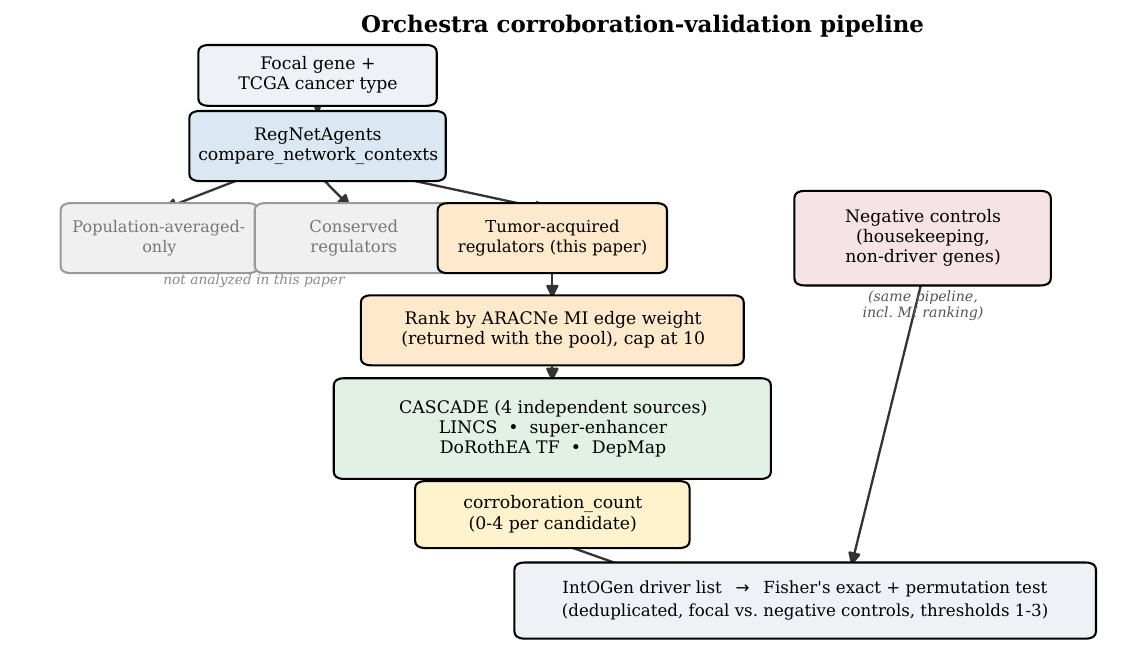}
\caption{The corroboration-validation pipeline. RegNetAgents' \texttt{compare\_network\_contexts} splits a focal gene's regulators into three disjoint tiers -- conserved (in both networks), population-averaged-only, and tumor-acquired (in the TCGA tumor network only). This paper analyzes only the tumor-acquired tier; the other two are shown for completeness. Tumor-acquired candidates (this paper's extension) are ranked by ARACNe MI edge weight and capped at 10, scored against four independent CASCADE evidence sources into a \texttt{corroboration\_count} (0--4), then compared against the IntOGen driver list via Fisher's exact and permutation tests, alongside an identical pipeline run on negative-control genes.}
\label{fig:pipeline}
\end{figure}

The ARACNe network itself, however, already assigns each edge a confidence value -- mutual-information edge weight -- and RegNetAgents (since v1.2.5) returns it alongside the pool, as a \texttt{tumor\_\allowbreak state\_\allowbreak only\_\allowbreak weights} gene-to-weight map (the same quantity \texttt{query\_network} reports as its \texttt{likelihood} field, $0$--$1$ per regulator). When \texttt{rank\_tumor\_acquired=True}, Orchestra sorts the pool by descending weight before applying the cap; if the map is absent it falls back to the unranked order, which did not occur in any experiment reported here. Figure~\ref{fig:pipeline} summarizes the full validation pipeline, from this candidate selection through corroboration scoring and the driver-list comparison.

\subsection{Corroboration scoring: four independent evidence sources}
\label{sec:corroboration}

For each candidate regulator, we query four CASCADE tools directly, chosen for producing structured, machine-checkable fields rather than requiring inference from narrative text:

\begin{itemize}
\item \textbf{LINCS knockdown} (\texttt{find\_expression\_regulators}): one call per focal gene retrieves the top-300 genes whose own knockdown most affects that focal gene's expression in LINCS L1000 \citep{subramanian2017lincs}; a candidate is scored positive if it appears in this set.
\item \textbf{Super-enhancer} (\texttt{check\_genes\_super\_enhancers}): one batch call per query checks all candidates against dbSUPER super-enhancer annotations \citep{khan2016dbsuper} across 102 cell types.
\item \textbf{DoRothEA TF confidence} (\texttt{validate\_tf\_classification}): one call per candidate checks whether it is a DoRothEA-confirmed transcription factor \citep{garciaalonso2019dorothea}.
\item \textbf{DepMap essentiality} (\texttt{get\_depmap\_essentiality}): one call per candidate checks CRISPR essentiality \citep{tsherniak2017depmap}; positive if pan-cancer essential (Chronos score \citep{dempster2021chronos} $<-0.5$ in more than half of cell lines) or mean Chronos score $<-0.5$. Each is computed across all 1,186 cell lines (30 lineages) of DepMap Public 25Q3, unweighted by lineage and not restricted to lines matched to the tumor type; a gene absent from DepMap is scored 0 (no hit) for this source.
\end{itemize}

A candidate's \emph{corroboration count} is the sum of positive sources (0--4). These four lightweight, structured-field tools give a more direct measurement of the underlying evidence than keyword-matching CASCADE's generated narrative text.

\subsection{Focal gene panel and negative controls}
\label{sec:panel}

We used RegNetAgents' focal-gene panel \citep{bird2026regnetagents}, for direct comparability with its reported driver-gene enrichment: 12 BRCA genes (\texttt{TP53, MYC, CTNNB1, CCND1, BRCA2, PIK3CA, PTEN, RB1, ERBB2, ESR1, GATA3, CDH1}) and 13 COAD genes (\texttt{TP53, MYC, CTNNB1, CCND1, KRAS, APC, SMAD4, BRAF, PIK3CA, PTEN, FBXW7, TCF7L2, RNF43}), each queried against the corresponding TCGA cancer-type network (25 gene--cancer-type queries). \texttt{BRCA1}, also on that panel, has no regulators in the TCGA BRCA network and so has no tumor-acquired tier to score; it is omitted. We also used RegNetAgents' negative-control panels: housekeeping genes (\texttt{ACTB, GAPDH, HPRT1, LDHA, TUBB}) and tumor-expressed non-driver genes (\texttt{FASN, PCNA, PKM, VIM}), each tested against both cancer types (18 queries). \texttt{PABPC1}, formerly in the non-driver panel, is excluded because IntOGen lists it as a driver. The two sets are pooled into a single negative-control group for all analyses, not tested separately. All queries specified \texttt{epithelial\_cell} as the cell type -- \texttt{compare\_network\_contexts}'s default, and the appropriate GREmLN reference for the epithelial-origin carcinomas BRCA, COAD, and STAD. This parameter selects only the GREmLN population-averaged network; the TCGA tumor-state network is set independently by the cancer type. This paper's analysis then operates entirely on the tumor-acquired regulators -- those in the TCGA network but not the GREmLN one.

These negative controls test a specific alternative explanation for any observed corroboration--driver association: three of the four CASCADE evidence sources (DoRothEA TF confidence, super-enhancer status, DepMap essentiality) are properties of the candidate regulator itself, independent of which gene it was queried against. If passing the $\geq 2$ threshold merely tracked how generally well-studied a candidate gene is -- rather than a genuine tumor-acquired regulatory relationship to a real cancer driver -- housekeeping and non-driver genes should show the same corroboration--driver enrichment as bona fide focal genes. Section~\ref{sec:main-results} tests this directly by applying the identical pipeline and thresholds to both groups.

\subsection{Ground truth: IntOGen}
\label{sec:groundtruth}

A candidate regulator was labelled a known cancer driver if its symbol appears in the IntOGen Compendium of Mutational Cancer Driver Genes \citep{martinezjimenez2020intogen} (release 2024.09.20; 633 genes, any cancer type and any role), the same snapshot that RegNetAgents and Orchestra's own driver annotation use; candidate selection is by ARACNe MI edge weight and never uses driver status, so sharing the file does not make the scoring circular. A second list, derived separately, serves as a robustness check in Section~\ref{sec:cgc-check}: the 125 mutation-driver genes of Table~S2A in \citet{vogelstein2013landscapes}. IntOGen calls drivers by positive selection in tumor somatic-mutation data, so it omits genes altered mainly by copy-number change or fusion, and both lists share a mutation-derived basis; IntOGen also draws in part on TCGA cohorts, the same data family as the tumor-state networks. These properties are discussed in the Limitations.

\subsection{Statistical framework}
\label{sec:stats}

Corroborated-vs-uncorroborated known-driver enrichment was assessed with Fisher's exact test (one-tailed, alternative = ``greater'') at three corroboration thresholds ($\geq 1$, $\geq 2$, $\geq 3$ of 4 sources), for the focal-gene group and, separately, the pooled negative-control group. Within each cancer-type panel -- BRCA/COAD (Table~\ref{tab:main}) and STAD (Table~\ref{tab:stad}) -- the three thresholds $\times$ two groups give up to six comparisons, and Benjamini--Hochberg FDR correction was applied across the testable ones as a per-panel correction family. A comparison at a given threshold is untestable if no candidate reaches it: with zero corroborated candidates, there is nothing to compare against the uncorroborated ones. All six were testable in BRCA/COAD; in STAD, no candidate had $\geq 3$ of the four sources positive in either the focal or the control group, so both $\geq 3$ comparisons drop, leaving four.

The primary analysis counts each (candidate regulator, focal gene) pair as one row, so a regulator appearing in several focal genes' tumor-acquired tiers contributes several rows. To check this does not inflate significance through pseudo-replication -- non-independent rows, violating Fisher's independence assumption -- we also collapsed the focal-gene set to one row per unique regulator (first occurrence) and repeated the test. The negative-control set is not collapsed; keeping all its rows gives that test more power to detect enrichment, so its null result is not an artifact of a smaller sample. As a further check not relying on Fisher's exact test's own assumptions, we additionally ran a 1,000-draw permutation test on the deduplicated set: for each draw, we randomly relabeled which genes were ``corroborated'' (preserving the true class sizes) and recomputed the corroborated-vs-uncorroborated known-driver rate gap; the empirical $p$-value is the fraction of permuted draws meeting or exceeding the observed gap.

Formal Benjamini--Hochberg correction is applied only to the primary comparisons: for each threshold ($\geq 1$, $\geq 2$, $\geq 3$) and each group (focal genes, negative controls), a Fisher's exact test of known-driver rate in corroborated versus uncorroborated candidates. That is 10 tests in total, 6 in BRCA/COAD and 4 in STAD (Table~\ref{tab:main}, Table~\ref{tab:stad}), corrected as a separate family per panel. The other tests that re-check the same $\geq 2$ finding -- the deduplicated and permutation re-tests, the exact cross-check reproducing RegNetAgents' published results, the second-driver-list re-tests (Section~\ref{sec:cgc-check}), and the stratified and logistic-regression tests of Section~\ref{sec:incremental-value} -- are reported individually rather than pooled with the primary comparisons. The MI-edge-weight analyses of Section~\ref{sec:ranking-check} and the per-source single-predictor comparisons of Section~\ref{sec:incremental-value} are descriptive and uncorrected.

\subsection{A second, separately constructed panel: STAD}
\label{sec:stad-methods}

To test whether the observed pattern generalizes beyond RegNetAgents' original panel, we built a second panel from a third cancer type, STAD. Some of its focal genes overlap with the BRCA/COAD panel -- both draw on common pan-cancer drivers -- but each gene's tumor-acquired regulator pool here comes from STAD's own ARACNe network, built from a different patient cohort, so the data are independent even where gene names coincide.

We assembled 15 candidate focal genes: four for which CASCADE's knockdown-effect predictions were validated against STAD patient data \citep{bird2026cascade} (\texttt{MYC, CCNE1, TOP2A, CCND3}); \texttt{AURKA}, whose predictions CASCADE validated in BRCA and COAD but did not test in STAD; and ten genes recurrently altered in gastric adenocarcinoma \citep{tcga2014gastric} (\texttt{TP53, ARID1A, PIK3CA, CDH1, RHOA, KRAS, SMAD4, ERBB2, APC, CTNNB1}).

Before committing to the panel, we pre-screened every gene for GREmLN population-averaged network coverage -- a lightweight \texttt{compare\_network\_contexts} pass without \texttt{validate\_tumor\_acquired}, which surfaces coverage gaps without incurring CASCADE call costs. One gene, \texttt{RHOA}, is absent from the GREmLN epithelial-cell network. Without a GREmLN regulator set for the focal gene, its tumor-acquired tier -- the TCGA regulators not also present in GREmLN -- is undefined, so it could not contribute any candidates to score. It was dropped at this stage, before any experimental run, so it is not counted as an in-experiment failure, leaving 14 focal genes.

The remaining 14 genes, plus the Section~\ref{sec:panel} negative-control set (9 genes, likewise pre-confirmed covered) tested against STAD, were carried into the full run: 23 queries (14 focal plus 9 control), zero coverage-related failures, again with \texttt{rank\_tumor\_acquired=True} selecting each query's 10 validated candidates by MI edge weight (Section~\ref{sec:ranking}).

\section{Results}

\subsection{In BRCA/COAD, a single evidence source is not diagnostic; multi-source agreement is}
\label{sec:main-results}

Table~\ref{tab:main} summarizes results across all three corroboration thresholds, for both the focal-gene and pooled negative-control groups, from 427 scored candidates (249 focal, 178 negative control), with the 10 tumor-acquired regulators per query selected by descending ARACNe MI edge weight (Section~\ref{sec:ranking}).

\begin{table}[h]
\centering
\caption{Corroboration threshold vs. known-driver (IntOGen) enrichment, BRCA/COAD panel (MI-weight-ranked candidate selection).}
\label{tab:main}
\begin{tabular}{llccccc}
\toprule
Group & Threshold & Corroborated & Uncorroborated & OR & Fisher $p$ & BH-adj. $p$ \\
\midrule
Focal genes & $\geq 1$/4 & 35/194 (18.0\%) & 7/55 (12.7\%) & 1.51 & 0.2383 & 0.7148 \\
Focal genes & $\geq 2$/4 & 15/46 (32.6\%) & 27/203 (13.3\%) & 3.15 & \textbf{0.0027} & \textbf{0.0162} \\
Focal genes & $\geq 3$/4 & 0/3 (0.0\%) & 42/246 (17.1\%) & 0.00 & 1.0000 & 1.0000 \\
Negative controls & $\geq 1$/4 & 4/135 (3.0\%) & 1/43 (2.3\%) & 1.28 & 0.6508 & 1.0000 \\
Negative controls & $\geq 2$/4 & 1/38 (2.6\%) & 4/140 (2.9\%) & 0.92 & 0.7037 & 1.0000 \\
Negative controls & $\geq 3$/4 & 0/2 (0.0\%) & 5/176 (2.8\%) & 0.00 & 1.0000 & 1.0000 \\
\bottomrule
\end{tabular}
\end{table}

\begin{figure}[H]
\centering
\includegraphics[width=\textwidth]{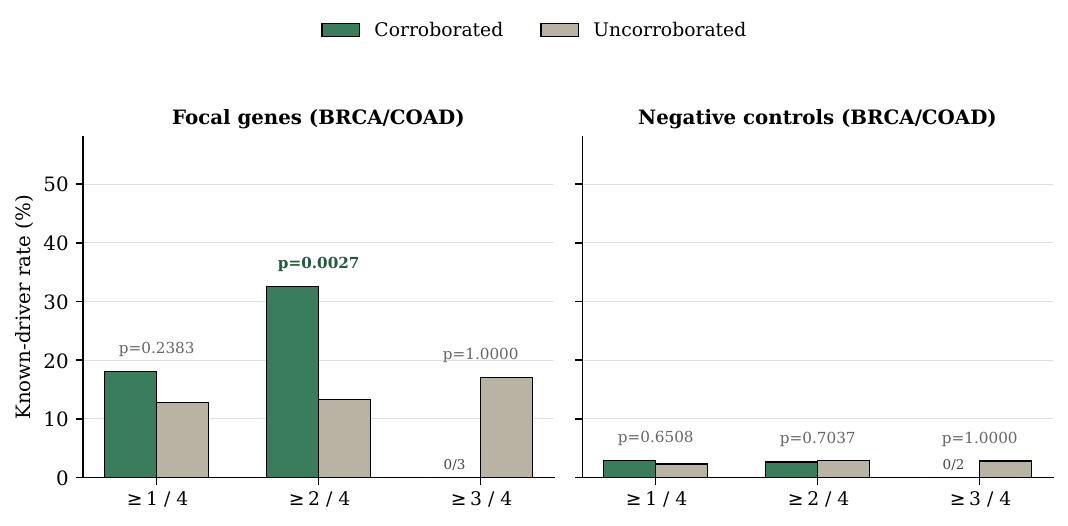}
\caption{Corroborated vs.\ uncorroborated known-driver rate by threshold (Table~\ref{tab:main}), focal genes (left) vs.\ negative controls (right); $p$-values are Fisher's exact (uncorrected), as in Table~\ref{tab:main}. Labels on empty bars give known drivers among corroborated candidates (0/3 focal, 0/2 control at $\geq 3$/4). At $\geq 1$/4, neither group shows a significant gap; at $\geq 2$/4, focal genes reach significance (bold $p$) while negative controls do not -- the pattern that motivates the $\geq 2$ threshold (Section~\ref{sec:threshold}).}
\label{fig:summary}
\end{figure}

At $\geq 1$, neither focal genes ($p=0.2383$) nor negative controls ($p=0.6508$) reach significance -- one plausible reason, that MI-weight selection already concentrates real candidates before any corroboration check is applied, is examined in Section~\ref{sec:ranking-check}. At $\geq 2$, the pattern that matters emerges cleanly: focal-gene significance appears where it was absent at $\geq 1$ ($p=0.0027$, BH-adjusted $p=0.0162$, the strongest result in the table) while negative controls show none ($p=0.7037$). At $\geq 3$ of 4 sources, only 3 focal-gene candidates and 2 negative-control candidates reach the threshold at all, and none of them is a known driver -- too few in either group to be informative. In this panel, corroboration separates focal genes from negative controls only at $\geq 2$. Known drivers are rare among negative-control candidates (5 of 178 rows), so the control comparisons have limited power to detect a modest association; their value is as a check against the large effect seen in focal genes, not as a precise estimate. Figure~\ref{fig:summary} plots the corroborated and uncorroborated known-driver rates by threshold for both groups.

\subsection{MI edge weight predicts candidate quality in BRCA/COAD, and shows at most a weak association with corroboration count}
\label{sec:ranking-check}

Because Table~\ref{tab:main}'s selection mechanism is itself the paper's contribution, we checked directly whether MI edge weight carries real signal in each panel's MI-ranked runs (389 focal candidate-rows: 249 BRCA/COAD, 140 STAD). Within each query's capped 10-candidate set, position matters in BRCA/COAD: the top 3 candidates by MI edge weight are known drivers at 29.3\% (22/75), versus 8.1\% (6/74) for the bottom 3 of the same 10 -- a large, significant gap (Fisher's exact $p=0.0008$) despite both groups being drawn from the identical, already-capped candidate pool. The effect is graded across the full ranking, not an artifact of the top-3/bottom-3 split: the Spearman correlation between MI-weight rank and known-driver membership across all 249 candidates is $\rho=-0.229$ ($p=0.0003$; negative because rank 1 denotes the highest weight). In STAD there is no such gradient: 14.3\% (6/42) versus 11.9\% (5/42), $p=0.50$ ($\rho=-0.045$, $p=0.60$, 140 candidates). Pooling the two panels gives 23.9\% (28/117) versus 9.5\% (11/116), $p=0.0025$, but that pooled figure is carried by BRCA/COAD and is not a property of both panels.

This selection may also explain why $\geq 1$ corroboration is not diagnostic in Table~\ref{tab:main}: MI-weight ranking already concentrates real candidates near the top of each query's 10, before any CASCADE evidence is checked, leaving less residual signal for a single source to add. We did not test this mediation directly (for example, comparing $\geq 1$ diagnosticity under MI-ranked versus unranked selection), so it is an interpretation of the Section~\ref{sec:main-results} pattern, not an independently established finding. The STAD panel fits this reading without testing it: there MI-weight ranking carries no detectable signal, and $\geq 1$ corroboration does discriminate focal genes (Section~\ref{sec:stad-results}).

Edge strength and corroboration count are at most weakly related. We measure edge strength by each candidate's rank within its query (rank 1 = strongest MI edge weight), so a \emph{negative} correlation means stronger edges are more corroborated.

Pooled across all 389 candidates there is no relationship ($\rho=-0.040$, $p=0.4293$): the top 3 and bottom 3 candidates by MI weight have similar mean corroboration counts (0.95 vs.\ 0.90) and reach the $\geq 2$ threshold at indistinguishable rates (Fisher's exact $p=0.51$). The pooled figures hide opposite directions in the two panels:
\begin{itemize}
\item \textbf{BRCA/COAD:} stronger edges are modestly more corroborated ($\rho=-0.131$, $p=0.0393$, uncorrected; mean count 1.07 for the top 3 vs.\ 0.84 for the bottom 3; $\geq 2$ in 15/75 vs.\ 10/74, Fisher's exact $p=0.2006$).
\item \textbf{STAD:} the direction is reversed and not significant ($\rho=+0.113$, $p=0.1829$; mean count 0.74 vs.\ 1.00; $\geq 2$ in 6/42 vs.\ 10/42).
\end{itemize}

A non-significant correlation does not prove independence: the test could miss a weak true effect, and in BRCA/COAD a weak one is nominally detectable. Still, the two scores are built from unrelated inputs (MI weight from the ARACNe co-expression network; corroboration count from four functional-evidence checks), and the visible association is weak and absent in STAD. The data therefore do not contradict the independence the paper's central claim depends on (Section~\ref{sec:intro}), though they do not establish it. Whether corroboration adds value with MI edge weight held fixed -- the question that matters for that claim -- is tested directly in Section~\ref{sec:incremental-value}.

\subsection{Robustness: deduplication}
\label{sec:dedup}

Of 249 focal-gene candidate-rows, 229 are unique genes; 17 genes repeat across multiple focal-gene queries (\texttt{TBC1D9} four times, \texttt{ZFP91} three times, and 15 genes twice each), so deduplication removes 20 rows (8\%). Collapsing to one row per unique gene (first occurrence) leaves the $\geq 2$ association essentially unchanged: OR $=2.92$, $p=0.0076$, versus OR $=3.15$, $p=0.0027$ on the raw candidate-rows (Benjamini--Hochberg-adjusted $p=0.0162$ within the BRCA/COAD family of 6 primary tests). The odds ratio moves by about 7\%; the $p$-value is larger because the sample is smaller, and, as for the other re-checks (Section~\ref{sec:stats}), it is reported individually rather than pooled into the corrected family. The 1,000-draw permutation test on the deduplicated set agrees (empirical $p=0.0050$, below). Because only 8\% of rows are removed, this is a check against exact duplicates, not a high-power test of pseudo-replication: the finding is robust to repeated candidates, and it does not address dependence between candidates arising from related focal-gene queries.

\subsection{Robustness: reproducing RegNetAgents' published results}

CTNNB1/BRCA's full tumor-acquired pool is exactly 10 genes, so MI-weight ranking reorders it without excluding anything: this experiment's candidate list (\texttt{DDR2, YAP1, IL6ST, VGLL4, CAPS, IRF7, ZFYVE28, ARID3A, NPAS2, OR4K5}, MI-weight order) is the same 10-gene pool RegNetAgents reports for this query, confirming the data pipeline reproduces their candidates rather than diverging from them. Two of the ten (\texttt{DDR2, IL6ST}) are IntOGen drivers; RegNetAgents' published table annotates driver status against a different driver list, so only the candidate pool is directly comparable.

\subsection{Robustness: permutation test}

On the deduplicated 229-gene set, 43 genes are corroborated ($\geq 2$) and 186 are not; 13/43 (30.2\%) of corroborated genes are known drivers versus 24/186 (12.9\%) of uncorroborated genes, an observed rate gap of 17.3 percentage points. We tested this gap with a 1,000-draw permutation test: in each draw, we randomly reshuffled which genes were labeled ``corroborated'' while holding the two group sizes fixed at 43 and 186, then recomputed the known-driver rate gap under that random relabeling. Only 5 of the 1,000 draws produced a gap at least as large as the one observed (empirical $p=0.0050$) -- consistent with the Fisher's-exact result, via a method that builds its null distribution directly from the data rather than from a distributional model.

\subsection{STAD: reproduction in a third cancer type}
\label{sec:stad-results}

Table~\ref{tab:stad} summarizes the STAD panel (Section~\ref{sec:stad-methods}), 230 scored candidates (140 focal from 14 genes, 90 negative control), with zero coverage-related failures, again with the 10 tumor-acquired regulators per query selected by descending ARACNe MI edge weight (Section~\ref{sec:ranking}).

\begin{table}[h]
\centering
\caption{Corroboration threshold vs. known-driver (IntOGen) enrichment, STAD panel (MI-weight-ranked candidate selection).}
\label{tab:stad}
\begin{tabular}{llccccc}
\toprule
Group & Threshold & Corroborated & Uncorroborated & OR & Fisher $p$ & BH-adj. $p$ \\
\midrule
Focal genes & $\geq 1$/4 & 17/93 (18.3\%) & 1/47 (2.1\%) & 10.29 & \textbf{0.0041} & \textbf{0.0083} \\
Focal genes & $\geq 2$/4 & 8/21 (38.1\%) & 10/119 (8.4\%) & 6.71 & \textbf{0.0012} & \textbf{0.0047} \\
Focal genes & $\geq 3$/4 & \multicolumn{5}{c}{insufficient data (0 corroborated)} \\
Negative controls & $\geq 1$/4 & 5/65 (7.7\%) & 0/25 (0.0\%) & $\infty$ & 0.1879 & 0.2506 \\
Negative controls & $\geq 2$/4 & 1/17 (5.9\%) & 4/73 (5.5\%) & 1.08 & 0.6582 & 0.6582 \\
Negative controls & $\geq 3$/4 & \multicolumn{5}{c}{insufficient data (0 corroborated)} \\
\bottomrule
\end{tabular}
\end{table}

No focal or negative-control candidate in STAD reaches the $\geq 3$ threshold at all (0 corroborated in either group), so the $\geq 3$ comparison is untestable for both groups in this panel. Unlike BRCA/COAD, $\geq 1$ is significant here in focal genes ($p=0.0041$; Benjamini--Hochberg-adjusted $p=0.0083$ within STAD's own testable-comparison family, Section~\ref{sec:stats}) but not in negative controls ($p=0.1879$; adjusted $p=0.2506$), so $\geq 1$ does discriminate focal genes from negative controls here -- unlike BRCA/COAD, where it discriminates neither (Section~\ref{sec:main-results}). At $\geq 2$, the diagnostic pattern that matters holds as well: focal genes reach significance (OR $=6.71$, $p=0.0012$ raw; deduplicating 140 candidate-rows to 135 unique genes gives OR $=6.40$, $p=0.0015$) while negative controls show no association ($p=0.6582$). As in BRCA/COAD, known drivers are rare among negative-control candidates (5 of 90 rows), so the control comparisons are best read as a check against the large focal-gene effect rather than a precise estimate.

\begin{figure}[H]
\centering
\includegraphics[width=0.85\textwidth]{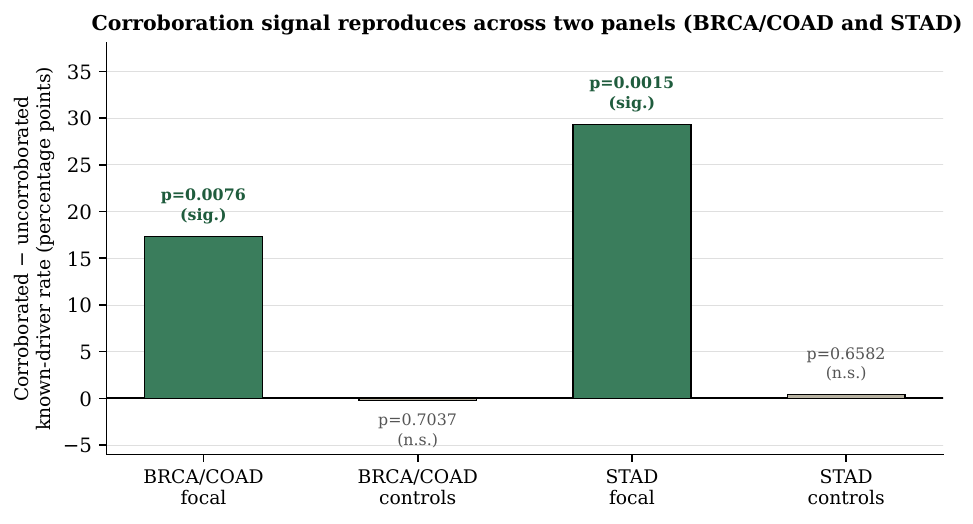}
\caption{Corroborated-vs-uncorroborated known-driver rate gap at the $\geq 2$ threshold, BRCA/COAD vs.\ STAD, deduplicated focal candidates (non-deduplicated negative controls, matching Section~\ref{sec:stats}'s methodology in both panels), MI-weight-ranked candidate selection throughout. Focal bars use the deduplicated $\geq 2$ results (Sections~\ref{sec:dedup}, \ref{sec:stad-results}); control bars are the $\geq 2$ rows of Table~\ref{tab:main} and Table~\ref{tab:stad}. Focal genes reach significance independently in both cancer types; negative controls do not, in either.}
\label{fig:stad}
\end{figure}

On the deduplicated STAD set, 21 genes are corroborated and 114 are not, an observed rate gap of 29.3 percentage points (8 of 21 versus 10 of 114 known drivers). The same permutation test as in BRCA/COAD -- reshuffling corroboration labels while holding the two group sizes fixed and recomputing the gap -- found only 2 of 1,000 draws meeting or exceeding this gap (empirical $p=0.0020$), agreeing with the Fisher's exact result as it did for BRCA/COAD. The effect size is larger than in BRCA/COAD (deduplicated OR $6.40$ vs.\ $2.92$), despite a smaller panel (135 vs.\ 229 deduplicated focal candidates). Figure~\ref{fig:stad} places the two panels side by side at the $\geq 2$ threshold: focal genes clear significance in both, negative controls in neither.

On the basis of two panels covering three cancer types -- BRCA and COAD from RegNetAgents' published panel, STAD newly constructed on a separate patient cohort -- we treat the $\geq 2$ pattern, under MI-weight candidate selection, as holding in the pooled BRCA/COAD panel and in STAD, not as a claim that it holds universally across all TCGA cancer types.

\subsection{Threshold justification}
\label{sec:threshold}

CASCADE's four evidence sources (Section~\ref{sec:corroboration}) are methodologically independent: transcriptional response to real knockdown perturbation, chromatin/epigenetic architecture, curated literature-and-motif-derived TF confidence, and functional CRISPR dependency. A single positive source means only that a candidate has some evidence in one of the four. Empirically, $\geq 1$ behaves differently across the two panels: in BRCA/COAD it separates neither focal genes nor negative controls from background at any correction level (Section~\ref{sec:main-results}); in STAD it is significant in focal genes after Benjamini--Hochberg correction ($p=0.0083$) but not in negative controls ($p=0.1879$; Section~\ref{sec:stad-results}). So $\geq 1$ is not uniformly uninformative the way $\geq 2$ is uniformly diagnostic.

We therefore do not rest the case for $\geq 2$ on the claim that $\geq 1$ fails everywhere -- which it does not. The threshold follows instead from what corroboration means: the corroboration count measures agreement among CASCADE's four independent sources, and agreement, by definition, requires at least two of them to be positive -- a single positive source cannot agree with itself. So $\geq 2$ is the smallest count at which multi-source agreement is possible, and we adopt it as the operational definition of agreement. Higher thresholds are not testable at current sample sizes ($\geq 3$ is reached by only 3 focal candidates in BRCA/COAD and none in STAD), so $\geq 2$ is also the strictest level we can evaluate.

The $\geq 2$ result also appears in STAD, an independent cohort, so it is not specific to BRCA/COAD.

Separately, whether this within-CASCADE agreement adds precision beyond RegNetAgents' own network-topology signal -- the cross-system question -- is tested in Section~\ref{sec:incremental-value}.

\subsection{Robustness: a second, separately derived driver list}
\label{sec:cgc-check}

Every result so far uses IntOGen as the sole ground truth (Section~\ref{sec:groundtruth}). To check that the $\geq 2$ pattern is not specific to one driver list, we repeated the analysis against a second, separately derived list: the 125 mutation-driver genes of Table~S2A in \citet{vogelstein2013landscapes}, assembled by a different group under its own criteria. The check is deliberately modest in scope: the list is small and flags few candidates, so it has little power either to confirm or to rule out an effect.

It does not reach significance in either panel, and we do not count it as support. In BRCA/COAD, corroborated focal candidates are Vogelstein drivers at 2 of 46 versus 4 of 203 for uncorroborated ones (OR $=2.26$, $p=0.3068$; deduplicated OR $=2.22$, $p=0.3143$; 1,000-draw permutation test, observed gap 2.5 percentage points, empirical $p=0.3280$). In STAD the corresponding counts are 1 of 21 versus 2 of 119 (OR $=2.92$, $p=0.3882$; deduplicated OR $=2.80$, $p=0.4003$; permutation gap 3.0 percentage points, empirical $p=0.4140$). The direction matches the IntOGen result in both panels, but the evidence is thin: only 6 of 249 BRCA/COAD and 3 of 140 STAD focal candidate rows are Vogelstein drivers at all, so the test has little power to detect an effect of any plausible size.

One result in this check runs against the intended pattern and is reported for completeness. In BRCA/COAD, 2 of 38 corroborated negative-control candidates are Vogelstein drivers against none of 140 uncorroborated ones (nominal $p=0.0446$; Benjamini--Hochberg-adjusted $p=0.2678$ across the panel's six comparisons). With two positives it does not survive correction and is not evidence of an effect, but it is the only place in the paper where a negative-control group is nominally significant. The reproduction script and its output are given in Section~\ref{sec:availability}.

\subsection{Does corroboration add value beyond the strongest single predictor?}
\label{sec:incremental-value}

In BRCA/COAD the strongest single predictor is not a CASCADE evidence source at all: MI edge weight (Section~\ref{sec:ranking-check}) separates its high- from low-ranked candidates on known-driver membership at $p=0.0008$, a lower $p$-value than any individual CASCADE source reaches there. In STAD it carries no detectable signal ($p=0.50$). Among the individual sources, DoRothEA TF confidence is the strongest on the deduplicated BRCA/COAD focal set ($p=0.0265$; no association in the non-deduplicated negative controls, $p=1.0000$), against $p=0.0076$ for the $\geq 2$ composite on the same set. The ordering differs in STAD, where super-enhancer status ($p=0.0075$) and DepMap essentiality ($p=0.0106$) are stronger single predictors than DoRothEA ($p=0.1169$); LINCS knockdown flags a single candidate in each panel and carries no weight. This raises a more stringent version of the question: does requiring genuine multi-source corroboration add anything once the single best predictor available -- MI edge weight, not just the best CASCADE source -- is already accounted for?

This section tests two related but separate questions, and they get different answers. The first, and the paper's central question, is whether corroboration adds value beyond MI edge weight itself; both panels agree that it does, strongly in STAD and more modestly in BRCA/COAD. The STAD test is the less demanding of the two, because MI edge weight carries no detectable signal there (Section~\ref{sec:ranking-check}); BRCA/COAD is the demanding test, and corroboration passes it modestly. The second, narrower question is whether corroboration adds value beyond DoRothEA specifically -- the strongest individual \emph{CASCADE} source in BRCA/COAD, though not in STAD, where that test is correspondingly less demanding; here the two panels disagree. We report both, since the second is a real and unresolved nuance about which evidence source carries the corroboration signal, even though it does not change the answer to the first.

Both questions are tested on the deduplicated focal set (MI-weight top-3 convention as in Section~\ref{sec:ranking-check}), by two instruments. The \textbf{regression} is primary: a logistic regression of known-driver membership on MI-top-3 (binary) and \texttt{corroboration\_count} (0--4), with a likelihood-ratio test against the MI-top-3-only model -- it uses every candidate and estimates corroboration's effect with MI edge weight held fixed statistically. The \textbf{stratified test} is a narrower cross-check that holds MI edge weight fixed \emph{by construction} instead: it discards every candidate in its query's MI-top-3 -- the slice where MI edge weight already does the work -- and asks whether, among the rank-4--10 remainder, $\geq 2$ corroboration still raises the known-driver rate (Fisher's exact plus a 1,000-draw permutation test). Both ask: once MI edge weight is accounted for, does corroboration still explain known-driver status?

\textbf{Corroboration beyond MI edge weight.} The regression is significant in both panels. The stratified cross-check is strongly significant in STAD but does not reach significance in BRCA/COAD, where few of the rank-4--10 candidates are corroborated.

In BRCA/COAD, adding \texttt{corroboration\_count} to the regression improves fit over the MI-top-3-only model: the likelihood-ratio test gives $p=0.0386$, and the \texttt{corroboration\_count} coefficient in that same model is significant on its own ($p=0.0409$). The stratified cross-check points the same way -- of the 165 candidates outside their query's MI-top-3, the 31 corroborated ones are known drivers at 19.4\% versus 9.0\% for the other 134 (OR $=2.44$) -- but does not clear significance ($p=0.0926$ Fisher, $p=0.0940$ permutation); with only 31 corroborated candidates the test has limited power, and the result is not evidence of a null effect. The evidence in this panel is therefore modest and rests on the regression.

In STAD, the effect is large: of the 94 candidates outside their query's MI-top-3, the 15 corroborated ones are known drivers at 40.0\% versus 7.6\% for the other 79 (OR $=8.11$), a gap the stratified cross-check finds strongly significant (Fisher $p=0.0032$; permutation $p=0.0030$). The regression agrees: \texttt{corroboration\_count} is highly significant ($p=0.0004$), while the MI-top-3 term is not once corroboration is in the model ($p=0.5694$); the likelihood-ratio test against the MI-top-3-only model gives $p=0.0001$.

Corroboration adds significant value beyond MI edge weight in both panels, strongly in STAD and modestly in BRCA/COAD. This is the most direct evidence in this paper for Orchestra's central architectural claim: cross-system corroboration captures real signal that MI edge weight, the strongest single predictor in BRCA/COAD, does not -- and in STAD, where MI edge weight is uninformative, corroboration is itself the dominant predictor.

\textbf{Corroboration beyond DoRothEA alone.} The same design, but substituting DoRothEA (the strongest single CASCADE source in BRCA/COAD) for MI-top-3, and the count of the \emph{other three} sources (LINCS, super-enhancer, DepMap; 0--3) for \texttt{corroboration\_count}. The question: once DoRothEA is accounted for, do the remaining three add anything? The \textbf{regression} (known-driver membership on DoRothEA plus the other-three count, full focal set, likelihood-ratio test against the DoRothEA-only model) is again primary; the \textbf{stratified test} is again the narrower cross-check, here restricted to DoRothEA-negative candidates -- the slice where DoRothEA contributes nothing, so any lift from having $\geq 2$ of the other three is attributable to those three alone. The two panels disagree.

In STAD, the other three do add value on both instruments. The regression finds their count a significant predictor ($p=0.0009$) that improves fit over the DoRothEA-only model (likelihood-ratio test $p=0.0002$). The stratified cross-check agrees: of the 120 DoRothEA-negative candidates, the 8 carrying $\geq 2$ of the other three are known drivers at 50.0\% versus 8.9\% for the other 112 (OR $=10.2$; Fisher $p=0.0064$, permutation $p=0.0070$).

In BRCA/COAD, the regression says the opposite: the other-three count is not significant once DoRothEA is included ($p=0.3192$), and adding it does not improve fit (likelihood-ratio test $p=0.3161$). The stratified cross-check is directionally consistent (28.6\% vs.\ 12.7\% known-driver rate among DoRothEA-negative candidates, OR $=2.75$) but underpowered at 14 such candidates with $\geq 2$ of the other three (Fisher $p=0.1096$, permutation $p=0.1220$); the null comes from the regression on the full set, which has more power than the stratified slice.

Which of CASCADE's four sources actually drives the corroboration effect is left open: DoRothEA appears to account for it alone in BRCA/COAD but not in STAD, consistent with the per-source results above (in STAD, super-enhancer status and DepMap essentiality are individually informative; in BRCA/COAD only DoRothEA is), and we do not resolve this. It bears on CASCADE's internal composition, not on the paper's central claim -- that corroboration adds value beyond the single best predictor, MI edge weight, which the preceding paragraphs found positive in both panels, modestly in BRCA/COAD.

\section{Limitations}

Throughout this paper, ``independent'' refers to data, curating organization, or inference method -- a separately derived driver list, an independently constructed panel -- not to developers. RegNetAgents, CASCADE, and Orchestra share a single author, disclosed in the Introduction; no claim of developer independence is intended anywhere.

Of CASCADE's four sources, only LINCS tests the candidate$\to$focal-gene relationship directly -- whether the candidate's knockdown moves the focal gene's expression. DoRothEA, super-enhancer, and DepMap are intrinsic-property checks on the candidate alone, so a candidate can score highly on those three with no evidence that it regulates the queried focal gene specifically. Nor were these four sources evaluated in CASCADE's own paper, whose results use only its perturbation-propagation engine \citep{bird2026cascade}; their use here is their first test as a corroboration filter. DepMap essentiality is in addition derived from cultured cancer cell lines pooled across all lineages, not from patient tumors or tumor-type-matched lines, so a regulator essential only in the relevant tissue could receive no DepMap hit.

A possible confound: corroborated candidates may just be better-studied genes, which a driver list may favour regardless: curated lists favour heavily published genes, and cohort-based lists favour genes in well-sampled tumor types. The crude form -- any well-studied gene showing the corroboration--driver link -- is addressed by the negative controls, in which corroboration shows no association with driver status at $\geq 2$ in either panel (Section~\ref{sec:panel}); the second driver list is too sparse to add to this (Section~\ref{sec:cgc-check}). The narrower form -- the signal riding on one source rather than genuine multi-source agreement -- is what Section~\ref{sec:incremental-value} tests: beyond DoRothEA alone, the other three sources add value in STAD ($p=0.0009$) but not detectably in BRCA/COAD (regression $p=0.3192$; the stratified check has only 14 qualifying candidates, $p=0.1096$). That corroboration adds value beyond MI edge weight -- the paper's central claim -- holds in both panels, though modestly in BRCA/COAD ($p=0.0386$); that it adds value beyond DoRothEA alone holds only in STAD.

Deduplication (Section~\ref{sec:dedup}) removes exact repeats but not dependence among candidates arising from related focal-gene queries, which may make the Fisher and permutation $p$-values somewhat too small; a test grouped by focal gene was not run.

The 10-candidate cap on tumor-acquired regulators validated per query still applies under MI-weight ranking; it excludes the majority of each query's tumor-acquired pool in most queries (Section~\ref{sec:ranking}). In BRCA/COAD, MI edge weight carries signal about driver status within the capped set (Section~\ref{sec:ranking-check}), which suggests ranking before the cap favours better candidates; we did not compare ranked and unranked selection directly, and no such gradient is detectable in STAD. A candidate excluded by the cap is never evaluated; raising or removing the cap was outside this paper's scope, though nothing in the selection mechanism (Section~\ref{sec:ranking}) prevents it.

The supported claim is scoped to BRCA, COAD, and STAD -- three cancer types, two independently constructed panels. This is broader than a single-panel result but still short of a claim that the pattern holds universally across all 14 TCGA cancer types RegNetAgents and CASCADE support; each additional cancer type validated strengthens the claim incrementally rather than settling it definitively, and we make no claim beyond the three tested here.

The ground truth and the network evidence share a data lineage. IntOGen calls drivers by positive selection in tumor somatic-mutation cohorts that include TCGA samples, the patient population from which the tumor-state ARACNe networks were inferred, so shared source cohorts could correlate a candidate's network position with its driver status for reasons unrelated to the regulatory claim under test. IntOGen also captures only mutational positive selection, so genes that act mainly through copy-number change or fusion are scored as non-drivers; this chiefly costs statistical power, but it means candidates acting through those mechanisms cannot be credited. By contrast, circularity through shared literature is weaker for the primary key: DoRothEA is partly literature-curated, but IntOGen's driver calls come from tumor mutation data rather than from the literature. The secondary Vogelstein list is curated and does not have this advantage.

The negative-control panels (Section~\ref{sec:panel}) provide the primary safeguard against these concerns: at $\geq 2$ they show no association in either cancer type while focal genes do in both. The second driver list (Section~\ref{sec:cgc-check}) does not add independent support, because it flags too few candidates to confirm or exclude an effect. Known drivers are also rare among negative-control candidates, so the safeguard is a check against the large focal-gene effect rather than a precise estimate. Together these reduce the concern without eliminating it.

Similarly, Section~\ref{sec:ranking-check}'s claim that MI edge weight and corroboration count are independent -- that they capture different information -- rests on correlation tests that are non-significant when pooled and in STAD. In BRCA/COAD there is a weak but nominally significant association ($\rho=-0.131$, $p=0.0393$, uncorrected), so the two measures are not fully independent there; elsewhere, a non-significant result is not proof of independence, and we did not check whether the data were enough to detect a small correlation.

\section{Data and Code Availability}
\label{sec:availability}

Orchestra is available at \url{https://github.com/jab57/Orchestra} under the MIT license. The \texttt{validate\_tumor\_acquired} and \texttt{rank\_tumor\_acquired} parameters described here are implemented in \texttt{orchestra\_langgraph\_workflow.py} (\texttt{\_run\_network\_comparison\_path}) and the accompanying MCP tool schema in \texttt{orchestra\_mcp\_server.py}. Setting \texttt{validate\_tumor\_acquired=true} on a network-comparison query surfaces each tumor-acquired regulator's source-level hits, its $\geq 2$ tier flag, and an explicit caveat that single-source hits are not established as informative; default output for existing callers is unchanged apart from a one-line hint that the option exists. The results reported here were produced with release v1.3.11; Orchestra is archived on Zenodo under a concept DOI that resolves to the latest release (10.5281/zenodo.21192968).

Reproducing any result in this paper requires both child MCP servers Orchestra composes (Section~\ref{sec:architecture}): RegNetAgents \citep{bird2026regnetagents}, available at \url{https://github.com/jab57/RegNetAgents} (all results reported here were produced with v1.3.0, Zenodo DOI 10.5281/zenodo.23215310; concept DOI 10.5281/zenodo.18500027; other versions may return different candidate sets), and CASCADE \citep{bird2026cascade}, available at \url{https://github.com/jab57/CASCADE} (v1.5.0, Zenodo DOI 10.5281/zenodo.23248601; concept DOI 10.5281/zenodo.18883012), which supplies the four corroboration evidence sources of Section~\ref{sec:corroboration}. Its DepMap evidence source reads \texttt{CRISPRGeneEffect.csv} from DepMap Public 25Q3 (Chronos gene effect scores; 1,186 models $\times$ 18,435 genes; SHA-256 \texttt{fa44b6a7cefe8748aaa2bef21bd49aa105f7ebfe819510da3022266d88f24d51}). The TCGA tumor-state ARACNe networks \citep{giorgi2026aracne} are distributed under CC BY-NC-ND 4.0, are not redistributed with this work, and are installed by RegNetAgents' own setup script.

Scripts for this paper and their full results are included in the repository under \texttt{scripts/} and \texttt{outputs/} respectively:
\begin{itemize}
\item \texttt{scripts/\allowbreak experiment\_\allowbreak corroboration\_\allowbreak ranked\_\allowbreak brca\_\allowbreak coad.py} -- collects the BRCA/COAD panel's MI-weight-ranked tumor-acquired candidates and their four CASCADE evidence sources (the scored-candidate data behind Table~\ref{tab:main})
\item \texttt{scripts/\allowbreak experiment\_\allowbreak corroboration\_\allowbreak ranked\_\allowbreak stad.py} -- the same for the independently constructed STAD panel (Table~\ref{tab:stad})
\item \texttt{scripts/\allowbreak gene\_\allowbreak match\_\allowbreak guard.py} -- asserts every tool response is genuinely about the gene requested (zero mismatches across all 66 queries)
\item \texttt{scripts/\allowbreak experiment\_\allowbreak threshold\_\allowbreak tables.py} -- the threshold-by-group enrichment statistics of Tables~\ref{tab:main} and~\ref{tab:stad} (Section~\ref{sec:stats})
\item \texttt{scripts/\allowbreak experiment\_\allowbreak cgc\_\allowbreak crosscheck.py} -- the deduplication and permutation re-tests, and the second-driver-list check of Section~\ref{sec:cgc-check}
\item \texttt{scripts/\allowbreak experiment\_\allowbreak mi\_\allowbreak rank\_\allowbreak check.py} -- the MI edge weight ranking check of Section~\ref{sec:ranking-check}
\item \texttt{scripts/\allowbreak experiment\_\allowbreak mi\_\allowbreak weight\_\allowbreak incremental\_\allowbreak value.py} -- tests whether corroboration adds predictive value beyond ARACNe MI edge weight (Section~\ref{sec:incremental-value})
\item \texttt{scripts/\allowbreak experiment\_\allowbreak dorothea\_\allowbreak incremental\_\allowbreak value.py} -- tests whether corroboration adds predictive value beyond DoRothEA alone, and reports each evidence source as a single predictor (Section~\ref{sec:incremental-value})
\end{itemize}
Every scoring script takes its answer key explicitly (\texttt{--truth intogen} or \texttt{--truth vogelstein}) and refuses to run if the snapshot's SHA-256 hash differs from the pinned value. The primary key is the IntOGen compendium \citep{martinezjimenez2020intogen}, release 2024.09.20 (633 genes; CC0), as shipped with RegNetAgents. The secondary key is the 125-gene Table~S2A of \citet{vogelstein2013landscapes}; its supplement is not redistributed here and must be obtained from the article. Regenerating the tumor-acquired candidate data these checks score requires RegNetAgents and CASCADE configured as in Section~\ref{sec:architecture}, then re-running the two \texttt{experiment\_\allowbreak corroboration\_\allowbreak ranked} data-collection scripts (the first two listed above), which drive the full Orchestra pipeline across all 66 queries. Each data set was scored once under the design fixed before collection. The run manifests in \texttt{outputs/v4/} record the exact commits used. Orchestra's workflow and server code at collection time is identical to v1.3.11; CASCADE was at its v1.5.0 release commit.

\section{AI Usage Disclosure}

Claude Code (Anthropic) was used to assist with software implementation, statistical analysis, and paper drafting. The author reviewed and approved all analyses and text and takes full responsibility for the methods and conclusions.

\section{Acknowledgements}

This work builds directly on RegNetAgents \citep{bird2026regnetagents} and CASCADE \citep{bird2026cascade}, and uses RegNetAgents' BRCA/COAD focal-gene and negative-control panels for direct comparability with its reported results. We acknowledge the GREmLN development team \citep{zhang2025gremln} at the Chan Zuckerberg Initiative AI for the pre-trained gene embeddings and population-averaged regulatory networks underlying RegNetAgents, and the CZ CELLxGENE Discover platform \citep{cellxgene2025}, from whose data those networks derive; F. M. Giorgi and M. J. Alvarez for the tumor-state ARACNe networks \citep{giorgi2026aracne}, distributed through the \texttt{aracne.networks} compilation \citep{aracne_networks}; the IntOGen team \citep{martinezjimenez2020intogen} for the open driver-gene compendium used as the primary ground truth; and the authors of \citet{vogelstein2013landscapes} for the secondary driver list. The results shown here are in whole or part based upon data generated by the TCGA Research Network: \url{https://www.cancer.gov/tcga}.

\section{Funding}

No funding was received for this work.

\section{Competing Interests}

The author is the founder of Bird AI Solutions. This work was conducted independently and received no funding.

\bibliographystyle{plainnat}

\end{document}